\documentclass[aps,prr,superscriptaddress,twocolumn,nofootinbib,floatfix,preprintnumbers]{revtex4-1}

\usepackage[english]{babel}
\usepackage[T1]{fontenc}
\usepackage[colorlinks=true,linkcolor=blue,citecolor=blue,urlcolor=blue]{hyperref}
\usepackage{bm,bbm,amssymb,amsmath}
\usepackage{graphicx}
\usepackage{multirow,dcolumn,booktabs,enumerate}
\usepackage{isotope}
\usepackage{physics}
\usepackage{slashed}
\usepackage{nicefrac}
\usepackage[capitalize]{cleveref}

\allowdisplaybreaks[1]

\newcommand{\fmiq}{\, \text{fm}^{-3}}
\newcommand{\kev}{\,\text{keV}}
\newcommand{\mev}{\, \text{MeV}}

\begin{document}

\preprint{LA-UR-23-32163}

\title{Neutron matter from local chiral effective field theory interactions at large cutoffs}

\author{Ingo~Tews}
\email[E-mail:~]{itews@lanl.gov}
\affiliation{Theoretical Division, Los Alamos National Laboratory, Los Alamos, New Mexico 87545, USA}

\author{Rahul~Somasundaram}
\affiliation{Theoretical Division, Los Alamos National Laboratory, Los Alamos, New Mexico 87545, USA}
\affiliation{Department of Physics, Syracuse University, Syracuse, New York 13244, USA}

\author{Diego~Lonardoni}
\thanks{Current affiliation:~XCP-2, Eulerian Codes Group, Los Alamos National Laboratory, Los Alamos, New Mexico 87545, USA}
\affiliation{Theoretical Division, Los Alamos National Laboratory, Los Alamos, New Mexico 87545, USA}

\author{Hannah~G{\"o}ttling}
\affiliation{Technische Universit\"at Darmstadt, Department of Physics, 64289 Darmstadt, Germany}
\affiliation{ExtreMe Matter Institute EMMI, GSI Helmholtzzentrum f\"ur Schwerionenforschung GmbH, 64291 Darmstadt, Germany}

\author{Rodric~Seutin}
\affiliation{Max-Planck-Institut f\"ur Kernphysik, Saupfercheckweg 1, 69117 Heidelberg, Germany}
\affiliation{Technische Universit\"at Darmstadt, Department of Physics, 64289 Darmstadt, Germany}
\affiliation{ExtreMe Matter Institute EMMI, GSI Helmholtzzentrum f\"ur Schwerionenforschung GmbH, 64291 Darmstadt, Germany}

\author{Joseph~Carlson}
\affiliation{Theoretical Division, Los Alamos National Laboratory, Los Alamos, New Mexico 87545, USA}

\author{Stefano Gandolfi}
\affiliation{Theoretical Division, Los Alamos National Laboratory, Los Alamos, New Mexico 87545, USA}

\author{Kai Hebeler}
\affiliation{Technische Universit\"at Darmstadt, Department of Physics, 64289 Darmstadt, Germany}
\affiliation{ExtreMe Matter Institute EMMI, GSI Helmholtzzentrum f\"ur Schwerionenforschung GmbH, 64291 Darmstadt, Germany}
\affiliation{Max-Planck-Institut f\"ur Kernphysik, Saupfercheckweg 1, 69117 Heidelberg, Germany}

\author{Achim Schwenk}
\affiliation{Technische Universit\"at Darmstadt, Department of Physics, 64289 Darmstadt, Germany}
\affiliation{ExtreMe Matter Institute EMMI, GSI Helmholtzzentrum f\"ur Schwerionenforschung GmbH, 64291 Darmstadt, Germany}
\affiliation{Max-Planck-Institut f\"ur Kernphysik, Saupfercheckweg 1, 69117 Heidelberg, Germany}

\begin{abstract}
Neutron matter is an important many-body system that provides valuable constraints for the equation of state (EOS) of neutron stars.
Neutron-matter calculations employing chiral effective field theory (EFT) interactions have been extensively used for this purpose.
Among the various many-body methods, quantum Monte Carlo (QMC) methods stand out due to their nonperturbative nature and the achievable precision.
However, QMC methods require local interactions as input, which leads to the appearance of stronger regulator artifacts compared to non-local interactions.
To circumvent this, we employ large-cutoff interactions derived within chiral EFT ($400 \mev \leq \Lambda_c \leq 700 \mev$) for studies of pure neutron matter.
These interactions have been adjusted to nucleon-nucleon scattering phase shifts, the triton binding energy, as well as the triton $\beta$-decay half-life. 
We find that regulator artifacts significantly decrease with increasing cutoff, leading to a significant reduction of uncertainties in the neutron-matter EOS.
We discuss implications for the symmetry energy and demonstrate how our new calculations lead to a reduction in the theoretical uncertainty of predicted neutron-star radii by up to 30\% for low-mass stars.
\end{abstract}

\maketitle

\section{Introduction}

In the past decade, exciting multimessenger data on neutron stars (NSs) invigorated the field of dense-matter physics. 
Observations of heavy pulsars~\cite{Demorest:2010,Antoniadis:2013pzd,Cromartie:2019kug,Fonseca:2021wxt}, gravitational-wave observations of NS mergers~\cite{LIGOScientific:2017a,LIGOScientific:2017b,TheLIGOScientific:2017qsa}, and x-ray pulse-profile modeling of rapidly rotating pulsars~\cite{Riley:2019yda,Miller:2019cac,Riley:2021pdl,Miller:2021qha,Choudhury:2024xbk} provided a wealth of new information.
Many studies analyzing these measurements additionally use input from nuclear-theory calculations of pure neutron or neutron-rich matter~\cite{Annala:2017llu,Capano:2019eae,Dietrich:2020lps,Essick:2020flb,Raaijmakers:2021uju,Essick:2021kjb,Annala:2021gom,Koehn:2024set,Rutherford:2024srk}.
Neutron matter is an important many-body system that provides valuable constraints for the equation of state (EOS) of NSs, and hence, high-fidelity calculations of neutron matter are an important ingredient for astrophysical data analyses.
State-of-the-art neutron-matter calculations are based on the combination of systematically improvable nuclear interactions from chiral effective field theory (EFT)~\cite{Epelbaum:2008ga,Machleidt:2011zz} and modern many-body methods that solve the nuclear many-body problem. 
One of the main benefits of chiral EFT is its capability to provide robust uncertainty estimates (see, e.g., Refs.~\cite{Epelbaum:2014efa,Drischler:2020hwi}).

Among the various many-body methods, quantum Monte Carlo (QMC) methods stand out due to their ability to provide virtually exact nonperturbative solutions to few- and many-body nuclear systems~\cite{Carlson:2014vla}.
They solve the Schr\"odinger equation by performing a diffusion in imaginary time to project out the ground state of a system starting from a given trial wave function. 
With current algorithmic developments, QMC methods are stochastically exact methods and typically reach a precision of a few percent~\cite{Lonardoni:2017hgs}. 
QMC requires local interactions as input, and in recent years, local chiral EFT interactions up to next-to-next-to-next-to-leading order (N$^3$LO) have been constructed specifically for QMC methods~\cite{Gezerlis:2013ipa,Gezerlis:2014zia,Piarulli:2014bda,Lynn:2015jua,Piarulli:2017dwd,Somasundaram:2023sup}. 
The combination of chiral interactions and QMC methods has led to new theoretical predictions for atomic nuclei~\cite{Lonardoni:2017hgs,Lonardoni:2018nob,Piarulli:2017dwd} and dense matter~\cite{Lynn:2015jua,Tews:2018kmu,Lonardoni:2019ypg}.

\begin{figure*}[t]
\includegraphics[trim=1cm 1cm 2cm 0cm,clip=,width=0.99\columnwidth]{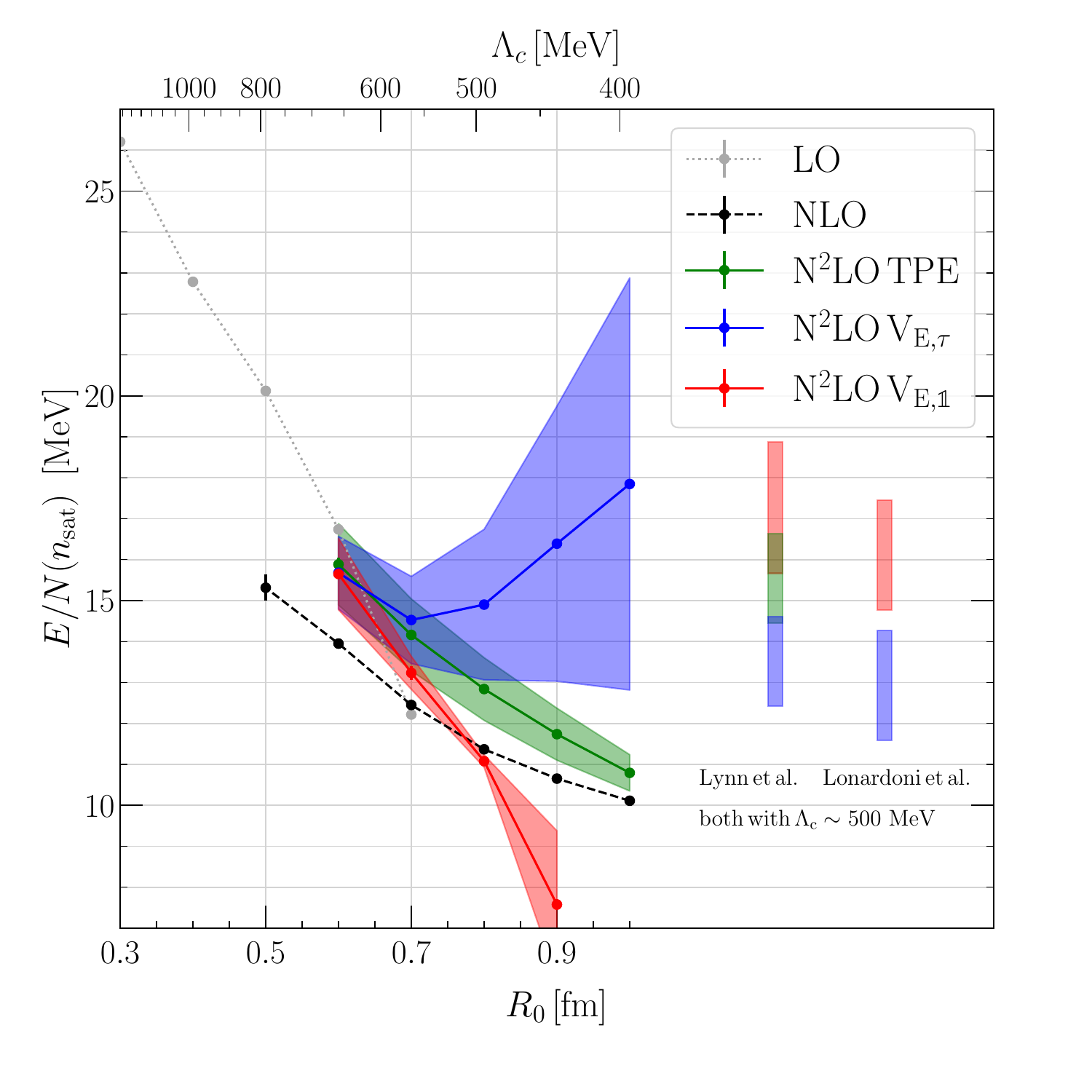} \hfill
\includegraphics[trim=1cm 1cm 2cm 0cm,clip=,width=0.99\columnwidth]{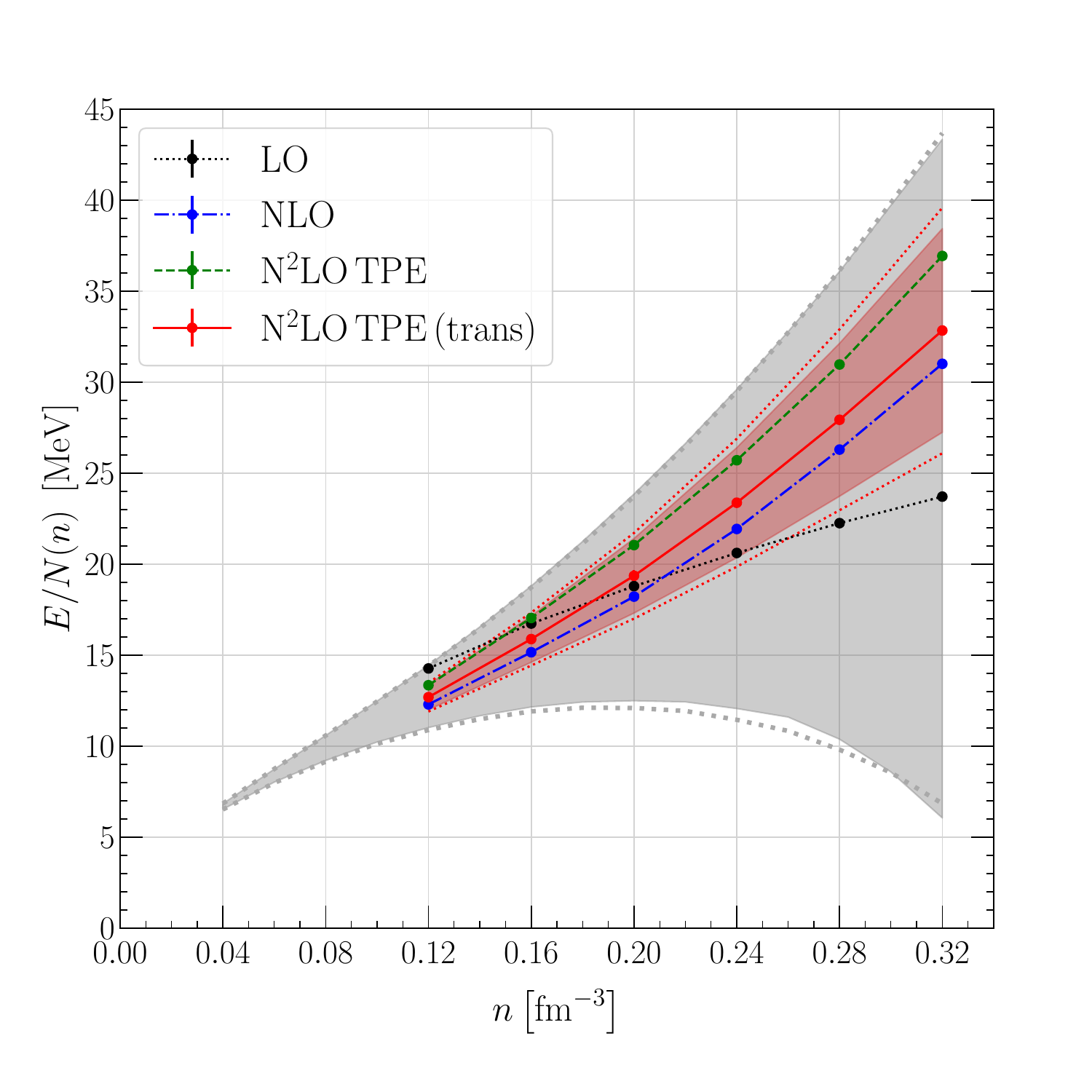}
\caption{Left panel: Neutron-matter energy per particle at saturation density as a function of coordinate-space cutoff $R_0$ and approximate momentum-space cutoff $\Lambda_c$ at leading order (LO), next-to-leading order (NLO), and next-to-next-to-leading order (N$^2$LO).
At N$^2$LO, we show the three different three-nucleon (3N) implementations of Ref.~\cite{Lynn:2015jua}.
Results are obtained from transient estimates and carry Epelbaum, Krebs, and Mei{\ss}ner (EKM) uncertainties.
We also compare with previous results of Lynn \textit{et al.}~\cite{Lynn:2015jua} [at diffusion Monte Carlo (DMC) level] and Lonardoni \textit{et al.}~\cite{Lonardoni:2019ypg} (transient estimates). 
These two calculations use the same computational setup as the present work but use a different local regulator function with $R_0=1.0$~fm, corresponding to $\Lambda_c\sim 500$~MeV.
Right panel: Neutron-matter energy per particle at different chiral effective field theory (EFT) orders as a function of density for $R_0=0.6$\,fm. 
At N$^2$LO, results are obtained by including only the 3N two-pion-exchange (TPE) interaction.
For our main results, we show the EKM uncertainty band and present uncertainties using the Gaussian process approach of Ref.~\cite{Drischler:2020hwi} (dotted red lines).
We compare with the lower-cutoff N$^2$LO results of Ref.~\cite{Lynn:2015jua,Tews:2018kmu} with EKM uncertainty (gray band) and using the Gaussian process approach (dotted gray lines).}
\label{fig:PNMresult}
\end{figure*}

However, microscopic calculations with nuclear interactions, such as those from chiral EFT, require regulators. 
The effects from the cutoff in the regulator can be absorbed by higher-order interactions, but may also lead to artifacts that can be sizable for local interactions~\cite{Lynn:2015jua}. 
Regulator effects are proportional to the inverse of the momentum-space cutoff but for local interactions lead to interactions in all partial waves, making it difficult to correct for regulator artifacts (see Refs.~\cite{Dyhdalo:2016ygz,Huth:2017wzw,Tews:2018sbi} for detailed discussions).
Local regulator artifacts can be especially impactful in neutron matter for typical cutoff choices~\cite{Lynn:2015jua,Tews:2018kmu}, increasing the uncertainty of the results and affecting astrophysical analyses.
Here, we expand on the work of Ref.~\cite{Tews:2018sbi} and employ local large-cutoff chiral interactions to next-to-next-to-leading order (N$^2$LO) with $400 \leq \Lambda_c \leq 700 \mev$, in studies of neutron matter.
We show our main results in Fig.~\ref{fig:PNMresult}.
We find that when increasing the momentum-space cutoff for local interactions, the impact of regulator artifacts reduces significantly, leading to a reduction of the uncertainty of the neutron-matter energy by a factor of $3$ compared with the lowest cutoff result.

\section{Local large-cutoff chiral interactions} 

The local chiral EFT Hamiltonians used here were constructed following the approach of Refs.~\cite{Gezerlis:2014zia,Somasundaram:2023sup}.
Up to N$^2$LO, contact operators are chosen to be local with the exception of the spin-orbit interactions, which can however be treated in QMC.
Regulators are chosen as specified in Ref.~\cite{Somasundaram:2023sup} and are Gaussian for short-range pieces. 
The interactions have been adjusted to neutron-proton scattering phase shifts by performing a least-squares minimization, details of which can be found in Ref.~\cite{Somasundaram:2023sup}.
The interactions used here span a range of coordinate-space cutoffs from $R_0 = 1.0$\,fm to $0.6$\,fm, approximately corresponding to $\Lambda_c \sim 400 \mev$ to $\sim 670 \mev$, respectively.
Typical cutoffs chosen in the community for calculations of neutron matter are in the range of $\sim 400-500$~MeV~\cite{Drischler:2017wtt,Keller:2022crb}.
This is because low-cutoff interactions show a better many-body convergence~\cite{Gezerlis:2014zia,Hoppe:2017lok}.
As QMC methods also show excellent convergence for high-cutoff interactions, here we can explore a larger range of cutoffs which allows us to reduce the size of regulator artifacts.
We are only limited in cutoff by the appearance of spurious bound states~\cite{Nogga:2005hy}.
Note that the least-squares fits here are slightly different than the ones reported in Ref.~\cite{Somasundaram:2023sup} because, here, the theoretical truncation uncertainty estimate is included in the cost function~\cite{Huth:2017wzw}.

In addition to the two-nucleon (NN) part of the Hamiltonian, at N$^2$LO the leading three-nucleon (3N) interactions enter.
In neutron matter, for nonlocal regulators only the parameter-free 3N two-pion-exchange (TPE) interaction contributes~\cite{Hebeler:2009iv} as the short-range 3N contact interaction ($V_E$) and the midrange one-pion-exchange--contact interaction ($V_D$) vanishes due to the Pauli principle and the pion coupling to spin.
With local regulators, however, the shorter-range terms contribute~\cite{Lynn:2015jua} because the local regulators induce a finite range when applied to contact interactions.
Effectively, this creates momentum-dependent 3N regulator artifacts that allow for three neutrons to interact.
The strength of these contributions is proportional to the respective low-energy couplings (LECs), $c_D$ and $c_E$, as well as the inverse of the cutoff $\Lambda_c^{-n}$, and depends on the chosen 3N operator structure~\cite{Huth:2017wzw}.
Hence, when studying local interactions in neutron matter, we need to include these shorter-range interactions and study their impact as a function of the cutoff scale.
Similarly to Ref.~\cite{Lynn:2015jua}, here we define two different 3N contact operators, $V_{E,\tau}$ and $V_{E,\mathbbm{1}}$ that we study in detail.
As the $V_{D}$ contribution is very small in neutron matter, even for low cutoffs $\Lambda_c$, we fix it to $V_{D,2}$~\cite{Lynn:2015jua}.

\begin{figure}[t]
\includegraphics[trim=1cm 1cm 2cm 2cm,clip=,width=0.95\columnwidth]{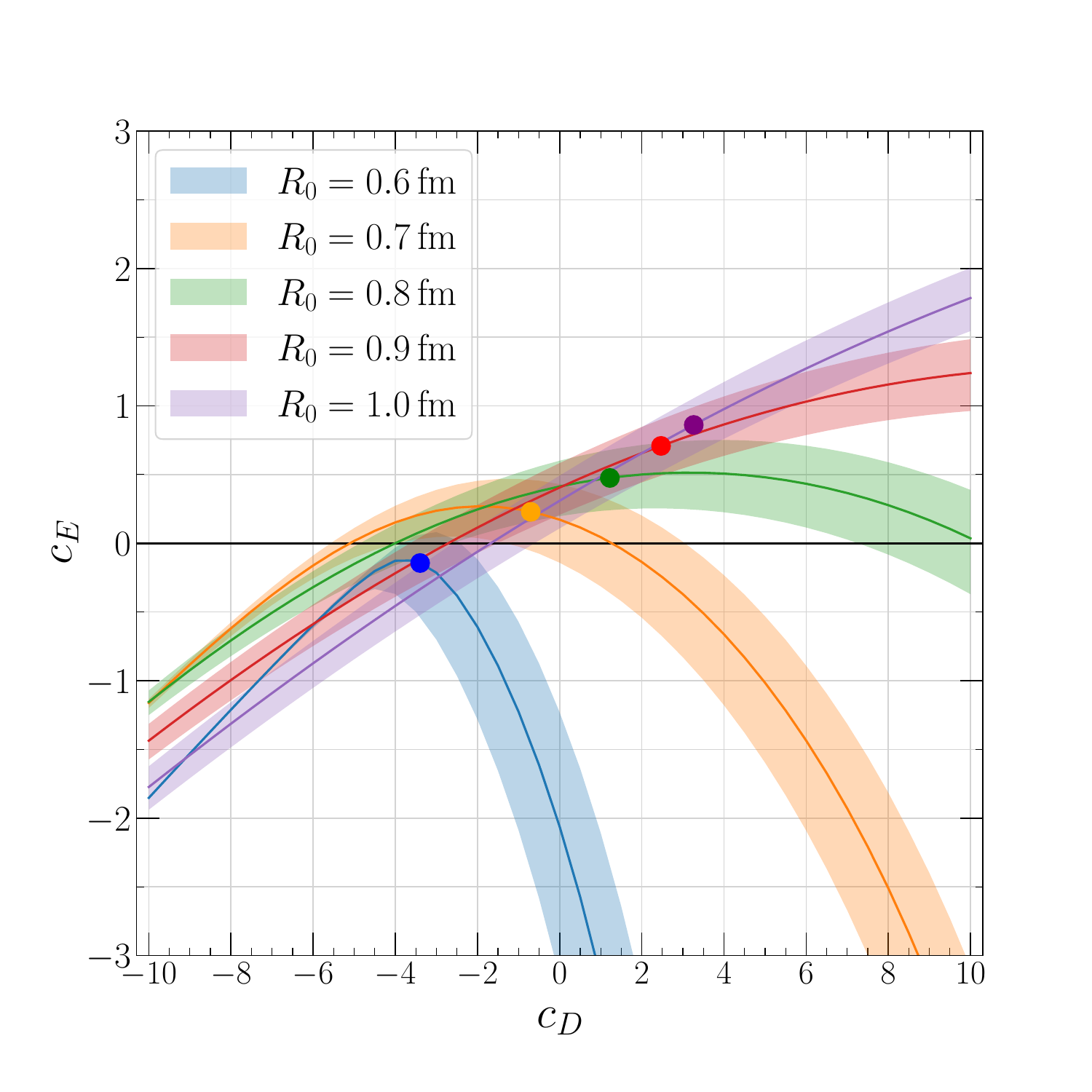}
\caption[]{Short-range low-energy coupling (LEC) $c_E$ as a function of $c_D$  for five different cutoff scales for the $V_{E,\tau}$ three-nucleon (3N) interaction. 
The bands are obtained by adjusting the 3N interactions to the $^3$H binding energy, assuming a 3\% uncertainty for the latter.
For each of the cutoffs, we have then adjusted the value of $c_D$ to the $^3$H $\beta$-decay half-life, indicated by the circles.}
\label{fig:3Nlec}
\end{figure}

To determine the unknown 3N LECs $c_D$ and $c_E$, we solve the Faddeev equations for the $^3$H binding energy for each operator choice and  cutoff (see, e.g., Ref.~\cite{Hebeler:2010xb}). 
This yields a relation between the two LECs for each cutoff, shown in Fig.~\ref{fig:3Nlec} for $V_{E,\tau}$, where we have assumed a 3\% theoretical uncertainty on the $^3$H binding energy for the fit.
Then for each cutoff, we adjust the LEC $c_D$ to the $^3$H $\beta$-decay Gamow-Teller matrix element (see also Refs.~\cite{Gazit:2008ma,Piarulli:2017dwd}).
This matrix element is sensitive to $c_D$ entering the axial-vector two-body currents.
Together, the $^3$H binding energy and Gamow-Teller matrix element allow us to determine both 3N LECs.
The resulting fit values are given in Table~\ref{tab:LECs}.
While the $c_D, c_E$ contributions only appear via regulator artifacts in pure neutron matter, the full 3N interactions as obtained here are necessary for future studies of systems containing protons such as nuclei and (a)symmetric nuclear matter.

\begin{table}
\centering
\tabcolsep=0.1cm
\def\arraystretch{1.2}
\begin{tabular}{c|cc|cc}
\hline\hline
& $V_{E,\tau}$ & & $V_{E,\mathbbm{1}}$ & \\ 
$R_0$~[fm] & $c_E$ & $c_D$ & $c_E$ & $c_D$ \\
\hline
0.6 & $-0.144$ & $-3.396$ & 0.134 & $-3.384$ \\
0.7 & 0.229 & $-0.704$ & $-0.207$ & $-0.699$ \\
0.8 & 0.476 & 1.223 & $-0.437$ & 1.240 \\
0.9 & 0.709 & 2.469 & $-0.665$ & 2.502 \\
1.0 & 0.861 & 3.264 & $-0.823$ & 3.304 \\
\hline\hline
\end{tabular}
\caption{Values for the short-range 3N LECs obtained from fits to the experimental $^3$H binding energy and $^3$H $\beta$-decay half-life (see Fig.~2).}
\label{tab:LECs}
\end{table}

\section{Results for pure neutron matter} 

To calculate the energy per particle of neutron matter we perform auxiliary-field diffusion Monte Carlo (AFDMC)~\cite{Schmidt:1999lik,Carlson:2014vla} calculations, similar to Refs.~\cite{Lynn:2015jua,Tews:2018kmu}.
Each simulation is performed in a finite box containing $66$ neutrons at a chosen density, which determines the box size. 
Starting from a trial wave function with plane-wave basis states, AFDMC recasts the problem into a diffusion equation and uses an imaginary-time evolution to project out the ground state, allowing us to extract the energy per particle.
To combat the sign problem, these simulations employ the constrained-path algorithm~\cite{Zhang:1995zz}, introducing a systematic uncertainty in the final result. 
For good trial wave functions, this systematic uncertainty is small, but for more complicated wave functions it can be of the order of a few MeV~\cite{Piarulli:2019pfq}.
To extract solutions, we release the constrained-path approximation in the end and perform fully unconstrained evolutions to extract the final result from the time evolution before stochastic noise begins to dominate~\cite{Lonardoni:2018nob}. 
Our final N$^2$LO results are based on such transient estimates.
More details on the method and the computational setup can be found in Ref.~\cite{Lynn:2019rdt}.

We begin by studying neutron matter at a fixed density, which we choose to be nuclear saturation density $n_{\rm sat}=0.16\,\textrm{fm}^{-3}$.
In the left panel of Fig.~\ref{fig:PNMresult}, we show the energy per particle at saturation density as a function of cutoff at leading order (LO), next-to-leading order (NLO), and N$^2$LO for the three different 3N-interaction implementations.
One Hamiltonian only includes the 3N TPE interaction, whereas the two other Hamiltonians additionally include the two different operator choices for $V_E$ mentioned before. 
The regulator artifacts are gauged by the difference between these three Hamiltonians: In the absence of artifacts, all three bands would coincide.
The values for $c_D$ and $c_E$ are fixed using the fitting procedure discussed above. 
At N$^2$LO, we give EFT truncation uncertainties estimated using the approach by Epelbaum, Krebs, and Mei{\ss}ner (EKM)~\cite{Epelbaum:2014efa} to compare with our previous results.
The EKM truncation uncertainty is based on analyzing results order by order.
Assuming the chiral EFT expansion to be a power series in $Q = \{m_{\pi},p\}/\Lambda_b$ with the typical momentum $p$ and the breakdown scale $\Lambda_b$, we can express any observable $X$ as a sum of all interaction terms:
\begin{equation}\label{eq:EFTexpansion}
X=X_0\sum_{k=0}^{\infty}c_k Q^{k}\,.
\end{equation}
In the EKM scheme, the error due to the truncation of the EFT expansion at $k=k_{\rm max}$ can be estimated as the first omitted term $X_0 c_{k_{\rm max}+1} Q^{k_{\rm max}+1}$.
The size of the unknown expansion coefficient $c_{k_{\rm max}+1}$ is estimated as the absolute maximum of the $k \leq k_{\rm max}$ coefficients. 
Note that we did not compute LO results for $R_0=0.8-1.0$~fm as we find that neutron matter collapses for these cutoffs~\cite{Tews:2015ufa}.
Consequently, the EKM uncertainty band at $R_0=0.8$~fm vanishes for the $V_{E,\mathbbm{1}}$ interactions as the NLO and N$^2$LO results are almost identical, leading to an almost vanishing expansion coefficient.
For large coordinate-space cutoffs (low momentum-space cutoffs), we expect regulator artifacts to be sizable as they are proportional to the inverse of the momentum-space cutoff to an even power, in our case, $n=2$. 
We find that the regulator artifacts lead to variations at the $\sim 5 \mev$ level and dominate over truncation uncertainties at the low cutoffs $\sim 400 \mev$.
When increasing the cutoff, we find that the regulator artifacts decrease as expected. 
At cutoffs $\sim 500 \mev$, we recover similar results to our previous calculations performed at a comparable cutoff~\cite{Lynn:2015jua,Lonardoni:2019ypg}, and regulator artifacts are of the same order of magnitude as the truncation uncertainties.
Cutoffs $\sim 500 \mev$ are the limit of what is commonly employed in the community.

In this work, we explore higher cutoffs, up to $\sim 700 \mev$.
We find that when increasing the cutoff to these values, the regulator artifacts decrease significantly and become much smaller than the EKM truncation uncertainty estimates. 
As a result, the energies for the three different 3N Hamiltonians at $R_0=0.6$\,fm agree remarkably well in the left panel of Fig.~\ref{fig:PNMresult}. 
At the constrained-path level, the 3N TPE and $V_{E,\mathbbm{1}}$ interactions are almost indistinguishable, with an energy difference of only $10 \kev$, and the $V_{E,\tau}$ interaction provides only an attractive contribution of $\sim 400 \kev$.
After the transient estimate, the difference reduces to $\sim 200\,\kev$, compared with the EKM uncertainty of $\sim 1 \mev$.
We have also checked the differences in the energy per particle for the different Hamiltonians at $2n_{\rm sat}$ at the constrained-path level.
The difference between 3N TPE and the $V_{E,\mathbbm{1}}$ interaction remains small and is $\sim 100 \kev$.
Similar to past work~\cite{Lynn:2015jua,Tews:2018kmu}, the attractive 3N regulator artifact ($V_{E,\tau}$) increases toward larger densities, and we find here an additional attraction of $\sim 2 \mev$. 
However, this is much smaller than the EKM truncation uncertainty of $\sim 5 \mev$ at $2n_{\rm sat}$.
We note that the reduction of the regulator artifacts is aided by the decreasing magnitude of $c_E$ resulting from our fits.
However, we stress that the increasing cutoff is the driving factor behind the reduction of the regulator artifacts. 
At $R_0=0.6$\,fm and $R_0=0.7$\,fm, the $c_E$'s are of similar size, see Table~\ref{tab:LECs}, but the regulator artifacts continue to decrease as the cutoff is increased, from $\sim900$~keV at $R_0=0.7$\,fm to $200$~kev at $R_0=0.6$\,fm. 
This is a result of regulator artifacts being proportional to inverse powers of $\Lambda_c$~\cite{Huth:2017wzw} and signals that the reduction of regulator artifacts is independent of the details of the 3N fit.

Finally, we note that the change of the energy per particle with cutoff in the left panel of Fig.~\ref{fig:PNMresult} is not flat at $R_0=1.0$~fm. 
This is due to the change of the NN interactions with cutoff, see, e.g., the NLO result as a comparison.
The NN interactions are adjusted to scattering phase shifts and lead to the best fit for $R_0=0.6$~fm~\cite{Somasundaram:2023sup}.
With lower $R_0$, the reproduction of NN scattering worsens.
Because of this, the vanishing regulator artifacts, and the appearance of spurious bound states for smaller $R_0$, we choose $R_0=0.6$ in the following.

\begin{figure*}[t]
\includegraphics[trim=0.7cm 0.5cm 0cm 0cm,clip=,width=0.7\columnwidth]{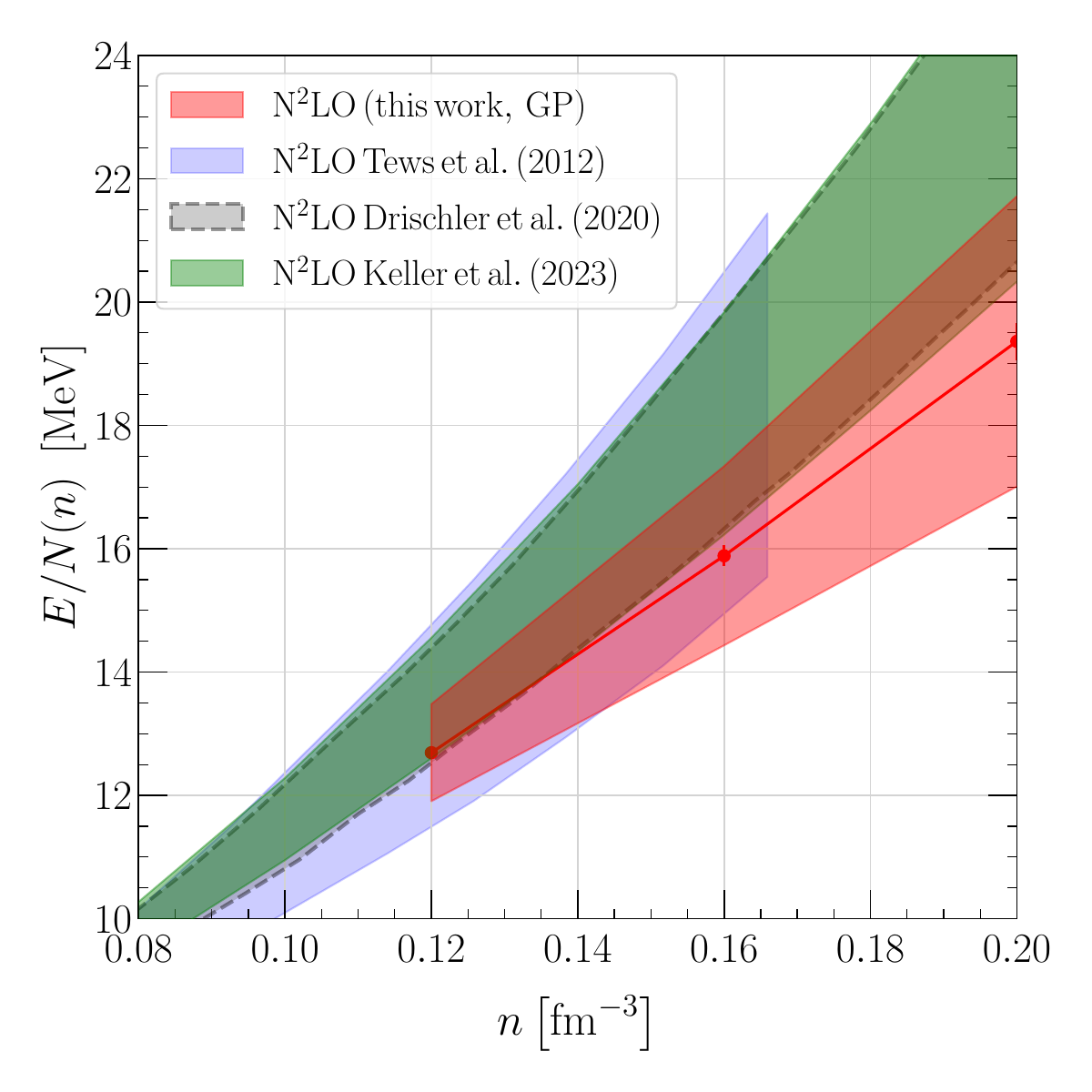}\hspace*{-0.25cm}
\includegraphics[trim=0.7cm 0.5cm 0cm 0cm,clip=,width=0.7\columnwidth]{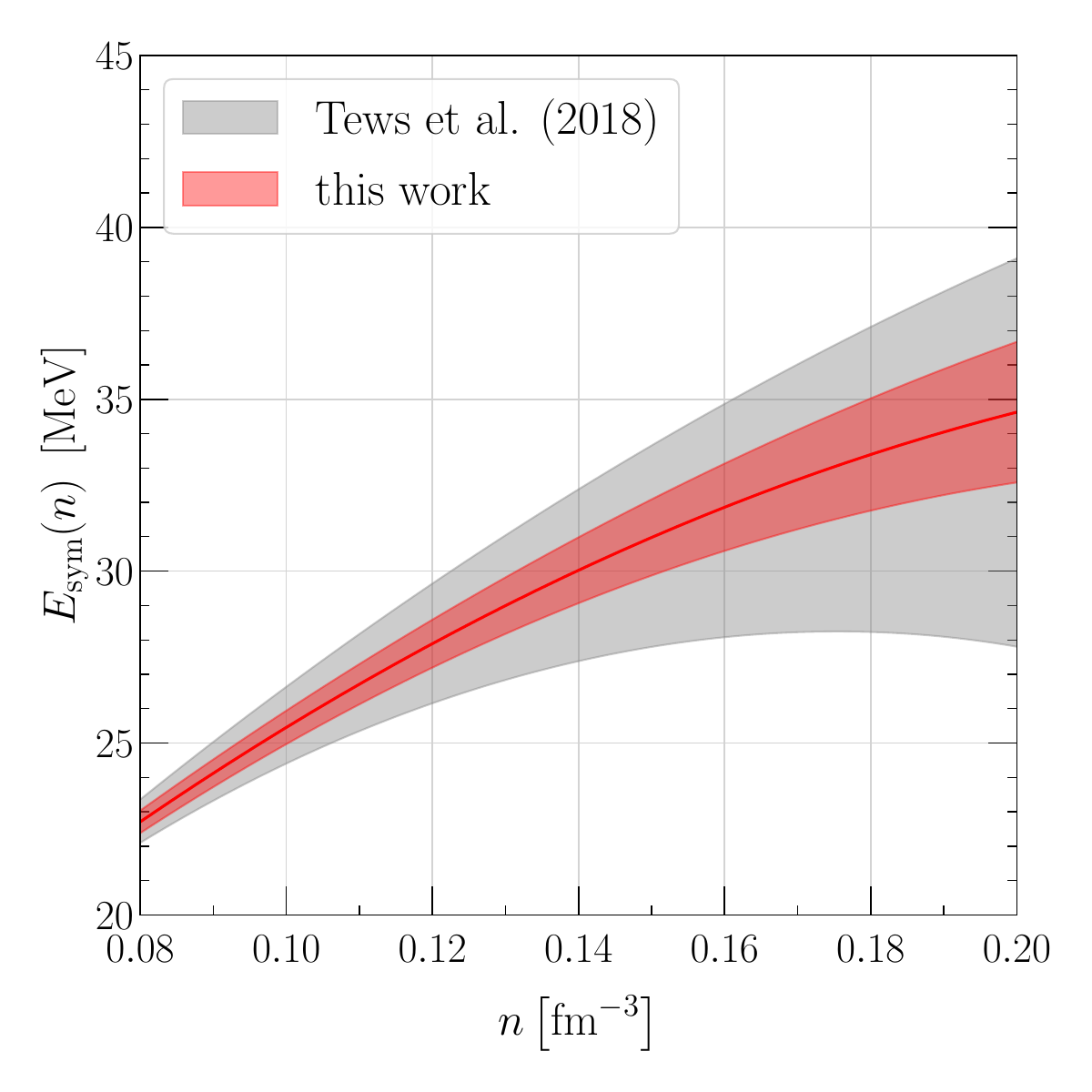}
\hspace*{-0.3cm}
\includegraphics[trim=0.7cm 0.5cm 0cm 0cm,clip=,width=0.7\columnwidth]{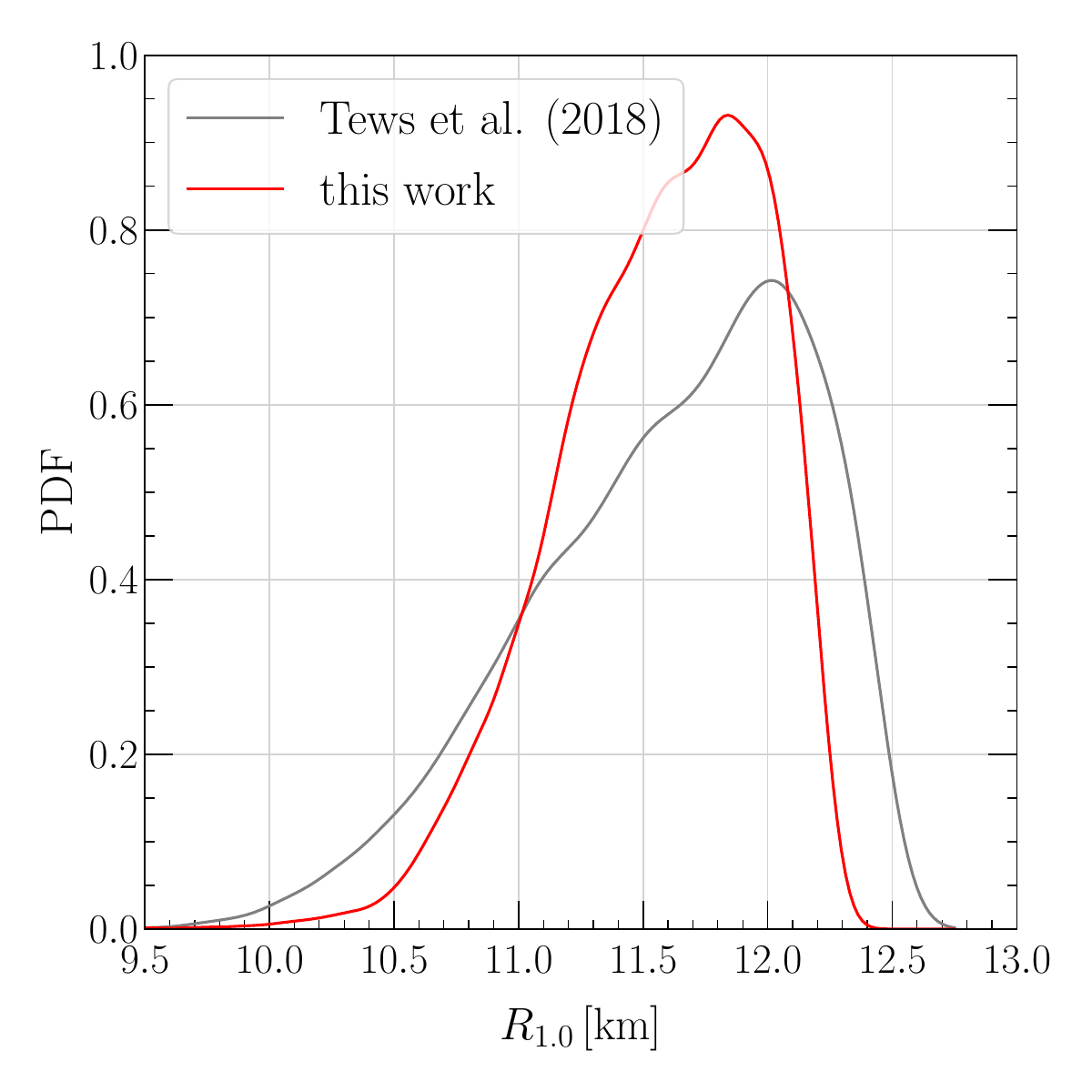}
\caption{Left panel: Comparison of different chiral effective field theory (EFT) calculations of the energy per particle of neutron matter at next-to-next-to-leading order (N$^2$LO). 
We show results of this work for $R_0=0.6$\,fm with Gaussian process (GP) Bayesian uncertainties and the N$^2$LO calculation of Tews \textit{et al.}~\cite{Tews:2012fj}. 
We also compare with Drischler \textit{et al.}~\cite{Drischler:2020yad} and Keller \textit{et al.}~\cite{Keller:2022crb} at a cutoff of $450 \mev$, both with GP Bayesian uncertainties.
Middle  panel: Symmetry energy as a function of density at N$^2$LO for $R_0=0.6$\,fm obtained from transient estimates and with EKM uncertainties.
We compare with previous results of Tews \textit{et al.}~\cite{Tews:2018kmu}. 
Right panel: Posterior probability distribution function (PDF) for the radius of a $1.0 \, M_{\rm sol}$ neutron star using the neutron matter results of this work as input compared with Tews \textit{et al.}~\cite{Tews:2018kmu} using  lower-cutoff N$^2$LO results.}
\label{fig:Esym_MR}
\end{figure*}

Having explored the cutoff dependence of the neutron-matter energy per particle, next we study its density dependence.
We focus on the neutron-matter EOS at N$^2$LO for $R_0=0.6$\,fm as it has minimal regulator artifacts.
Because regulator artifacts are small at this cutoff, in the following we employ only the 3N TPE interaction, which is the only contribution of the leading 3N forces to neutron matter for nonlocal regulators~\cite{Hebeler:2009iv}.
We show the results for the EOS in the right panel of Fig.~\ref{fig:PNMresult} at different chiral orders and, at N$^2$LO, at the constrained-path level and with transient estimate. 
The latter is our final result, and we have calculated uncertainties using both the EKM approach and the Gaussian process (GP) Bayesian uncertainty model from Ref.~\cite{Drischler:2020hwi}.
For the GP Bayesian uncertainties, we use a reference energy $E_{\rm ref}/N = 16\mev (n/n_{\rm sat})^{2/3}$, an expansion parameter $Q/\Lambda_b = k_{\rm F}/(600 \mev)$, and an inverse $\chi^2$ prior.
We find that both uncertainty estimates are consistent and our results are in excellent agreement with our previous N$^2$LO QMC calculations~\cite{Lynn:2015jua,Tews:2018kmu} but with significantly reduced uncertainties by a factor of $3$.
In the left panel of Fig.~\ref{fig:Esym_MR}, we compare our final result with the N$^2$LO calculations of Refs.~\cite{Tews:2012fj,Drischler:2020yad,Keller:2022crb}.
Our results are overall in good agreement but predict a slightly softer neutron-matter EOS.

\section{Impact on neutron stars}

Next, we study the impact of our calculations on the symmetry energy $E_{\rm sym}(n)$ and on NSs.
For this, we employ the metamodel of Refs.~\cite{Margueron2018a,Margueron2018b}, see also Ref.~\cite{Somasundaram:2020chb}.
The isovector parameters of the metamodel, i.e., the nuclear empirical parameters that govern the symmetry energy, are fit to the neutron-matter results presented above. 
We do not explore variations in the isoscalar metamodel parameters that determine the symmetric-matter EOS and fix these parameters at $E_{\rm sat}/A = -16$\,MeV, $n_{\rm sat} = 0.16 \fmiq$, $K_{\rm sat} = 230$\,MeV, and neglect the higher-order parameters. 
This is justified because the lower-order parameters are better constrained by nuclear experiments~\cite{Margueron2018a,Margueron2018b,Grams:2022bbq} and therefore, within uncertainties, are less important for $E_{\rm sym}(n)$ than the neutron-matter behavior. 
On the other hand, the higher-order isoscalar parameters play an important role but only at densities outside the range considered here. Therefore, while we have effectively fixed the EOS of symmetric matter, our calculation still provides a useful estimate of $E_{\rm sym}(n)$,
which is shown in the middle panel of Fig.~\ref{fig:Esym_MR}.
We find that our results predict $E_{\rm sym}(n_{\rm sat})=31.9\pm 1.3 \mev$ and its slope $L=\left. 3 n_{\rm sat} \frac{\partial E_{\rm sym}(n)}{\partial n}\right|_{n_{\rm sat}}=40.4\pm 8.1 \mev$.
These results are consistent with previous determinations from chiral EFT~\cite{Hebeler:2009iv,Tews:2012fj,Drischler:2020hwi} and the unitary-gas bound of Ref.~\cite{Tews:2016jhi}.
As common for chiral EFT predictions of $L$, the value is smaller than the one extracted from the Lead Radius Experiment (PREX)~\cite{Adhikari:2021pxi}.

Finally, we study the impact of our results on NS structure.
Following Ref.~\cite{Koehn:2024set}, we have generated 10,000 samples from the metamodel by sampling over its symmetry energy parameters in a uniform range $E_{\rm sym} = [28,39]$\,MeV , $L = [20,60]$\,MeV, and $K_{\rm sym} = [-300,0]$\,MeV. 
Higher-order empirical parameters are neglected and the isoscalar parameters are fixed as before.
Then we impose our N$^2$LO QMC results for neutron matter by ensuring that the EOS of each sample lies within the EKM uncertainty estimate shown in the right panel of Fig.~\ref{fig:PNMresult} up to $2 n_{\rm sat}$. 
Above $2 n_{\rm sat}$, we perform a general extrapolation in the speed of sound, see Ref.~\cite{Koehn:2024set} for more details. 
As in Ref.~\cite{Koehn:2024set}, the EOS of the crust is fixed to be the one from Ref.~\cite{Douchin:2001sv}.
The resulting NS EOS is used to calculate the structure of NSs by solving the Tolman-Oppenheimer-Volkoff (TOV) equations.

We show the resulting posterior for the radius of a $1.0 M_{\rm sol}$ NS $R_{1.0}$ in the right panel of Fig.~\ref{fig:Esym_MR}.
We focus here on light NSs, because their central densities are lower so that they are most sensitive to improvements in the EOS at nuclear densities.
Based only on the neutron-matter calculations and the existence of $2.0 M_{\rm sol}$ NSs, we extract a radius of $R_{1.0}=11.62^{+0.38}_{-0.46}$\,km ($R_{1.0}=11.70^{+0.48}_{-0.68}$\,km) for the new (previous) QMC neutron-matter result at the 68\% confidence level. 
For a typical $1.4 M_{\rm sol}$ NS, we find $R_{1.4}=11.70^{+0.38}_{-0.51}$\,km ($R_{1.4}=11.76^{+0.45}_{-0.62}$\,km) for the new (previous) QMC neutron-matter result.
We therefore find a 20-30\% reduction in the uncertainty of astrophysical NS observables based on the improved nuclear theory input. 

In summary, we have presented AFDMC calculations of the neutron-matter EOS with local chiral interactions at large cutoffs.
While local interactions can lead to sizable regulator artifacts in the EOS, we have shown that these regulator artifacts systematically decrease with increasing cutoff values and become smaller than the EFT truncation uncertainties for cutoffs of $\sim 700 \mev$. 
This leads to significantly reduced uncertainties compared with previous QMC calculations and provides improved constraints that can be employed in astrophysical studies of NSs and their mergers~\cite{Pang:2022rzc,Essick:2021kjb,Annala:2021gom,Rutherford:2024srk}.

\acknowledgements
We thank L.~Huth and J.E.~Lynn for insightful discussions.
I.T., S.G., and J.C.~were supported by the U.S. Department of Energy (DOE), Office of Science, Office of Nuclear Physics, under Contract No.~DE-AC52-06NA25396,
and by the DOE, Office of Science, Office of Advanced Scientific Computing Research, Scientific Discovery through the Advanced Computing (SciDAC) NUCLEI program.
I.T.~was also supported by the Laboratory Directed Research and Development program of Los Alamos National Laboratory under Project No. 20220541ECR. 
Ra.S.~acknowledges support from the Nuclear Physics from Multi-Messenger Mergers (NP3M) Focused Research Hub which is funded by the National Science Foundation under Grant No. 21-16686, and by the Laboratory Directed Research and Development program of Los Alamos National Laboratory under Project No. 20220541ECR.
The work of D.L.~was supported by the DOE NUCLEI SciDAC Program.
The work of H.G., K.H., and A.S.~was supported in part by the European Research Council (ERC) under the European Union's Horizon 2020 research and innovation programme (Grant Agreement No.~101020842) and by the State of Hesse within the Research Cluster ELEMENTS (Project ID 500/10.006). 
Ro.S.~was supported by the Max Planck Society.
Computational resources have been provided  by the Los Alamos National Laboratory Institutional Computing Program, which is supported by the DOE National Nuclear Security Administration under Contract No.~89233218CNA000001, and by the National Energy Research Scientific Computing Center (NERSC), which is supported by the DOE, Office of Science, under Contract No.~DE-AC02-05CH11231.

\bibliographystyle{plain}
\bibliography{biblio}

\end{document}